# Improving Electrical Contact Quality and Extraordinary Magnetoresistance in High Mobility III-V Semiconductors

Thierry Désiré Pomar[1], Tristan Steegemans[1], Sreejith Kumar[1], Rasmus Bjørk[1], Zijin Lei[3,4], Erik Cheah[3,4], Rüdiger Schott[3,4], Peter Bøggild[2], Nini Pryds[1], Werner Wegscheider[3,4] and Dennis Valbjørn Christensen[1,5]

[1]Department of Energy Conversion and Storage, Technical University of Denmark
[2]Department of Physics, Technical University of Denmark
[3]Solid State Physics Laboratory, ETH Zürich
[4]Switzerland Quantum Center, ETH Zürich
[5]Institute for Advanced Study, Technical University of Munich

October 23, 2024

## Abstract

Magnetometers based on the extraordinary magnetoresistance (EMR) effect are promising for applications which demand high sensitivity combined with room temperature operation but their application for magnetic field sensing requires further optimization. A key challenge is to obtain Ohmic metal/semiconductor contacts with low contact resistances in EMR devices comprising semiconductors with low carrier densities and high electron mobilities, yet, this topic remains scarcely investigated experimentally. By annealing high-mobility InSb in argon with systematically increasing temperatures, we experimentally demonstrate how the contact resistance to InSb films can be improved by two orders of magnitude by annealing to $3 \cdot 10^{-6} \Omega cm^2$ without degrading the high mobility. We further show that lowering the contact resistance monotonously increases the room temperature magnetoresistance at 2 T from 700% to 65,000%. Lastly, we explore the origin of intrinsic magnetoresistance in high-mobility InSb thin films and suggest that it can best be explained by multiple band conduction.

## 1 Introduction

Magnetoresistance is a property exhibited by some single-component or hybrid material systems where the measured electrical resistance varies as a result of an applied magnetic field. The different mechanisms which can produce magnetoresistance can be grouped into intrinsic effects related to the intrinsic properties of particular materials and geometric effects arising due to the way charge carriers interact with the device geometry. Extraordinary magnetoresistance (EMR) is a geometric magnetoresistance effect that can be observed in hybrid systems where one constituent material has a high carrier mobility and the second has a high conductivity.[1] When a voltage is applied across such a system, the resulting internal electrical field is perpendicular to the interface between the two materials, as the highly conductive phase is essentially an equipotential surface. In the absence of an external magnetic field, charge carriers follow the electrical field lines and move primarily through the highly conductive material. If an out-of-plane magnetic field is applied, however, the charge carriers near the interface are deflected due to the Lorentz force and are forced to travel through the high-mobility phase instead, with the angle of the deflection depending on product of the mobility, $\mu$, and the magnetic flux density, $B$. The resulting electrical resistance is therefore highly dependent on the magnetic field strength. EMR devices made with high mobility semiconductors interfaced with gold have shown room temperature magnetoresistance values as high as 750,000% at 8 T.[2].

The III-V semiconductor InSb is an obvious choice for EMR devices[3] as it can be grown with high quality at wafer scale and possesses the highest room temperature electron mobility of any bulk material.[4] However, numerical models suggest that the magnetoresistance depends strongly on the quality of the contact between the metal and semiconductor, where a high contact resistivity results in a total quenching of the EMR effect.[5–10]. For most III-V semiconductors placed in contact with a metal, the Fermi level at the surface of the semiconductor becomes pinned inside the band-gap, creating a Schottky barrier which electrons must overcome to transfer across the metal/semiconductor interface.[11] Annealing has been shown to produce an improved contact quality in III-V materials, including InSb[12], InGaAs[13, 14], and GaSb[15]. The decrease in the contact resistivity after annealing can arise if metal ions from the contacts diffuse into the semiconductor and form alloys.[14] These alloys can reduce the contact resistivity by increasing the carrier density in the semiconductor and therefore reducing the work function difference between the two materials[16], by providing a mechanism for trap-assisted tunneling [14], or by reducing the width of the Schottky barrier.

Most of the previously published literature on how the contact resistance influences the EMR effect has been based on finite element simulations.[5–9] In this work we experimentally examine the effect of thermal annealing on the contact resistance between InSb and Ti/Au layers and detail how the systematic decrease of the contact resistivity monotonously improves the room temperature magnetoresistance from 700% to 65,000%. In addition, we report how the material properties and intrinsic magnetoresistance evolve with increasing annealing temperatures and identify a thermal window in which contact resistivity can be reduced by two orders of magnitude without degradation of the semiconductor material properties.

## 2 Methods

Molecular beam epitaxy was used to grow the thin film heterostructures on GaAs substrates. Growing high quality InSb films via homo-epitaxy on InSb substrates is undesirable due to the substrate cost and the high conductivity of the material which allows for parallel conduction channels.[3] Semi-insulating GaAs substrates are typically used instead but the large lattice mismatch (14.6%) between the two materials creates defects including threading dislocations, micro-twins, and cracks.[17] Electron scattering can occur at these defect sites which is detrimental to the carrier mobility of the material near the GaAs interface. By growing thicker films, the effect of defects resulting from lattice mismatch can be mitigated as the lattice is allowed to relax away from the interface which reduces the defect density.[3] For EMR devices and other types of magnetometers such as Hall sensors, a strong signal response is achieved when the sheet carrier density is low, which makes overly thick films undesirable.[8] Thus, a tradeoff exists between making a film which is thick enough to display a high carrier mobility but thin enough that the sheet carrier density remains low. To further mitigate the lattice mismatch between the GaAs wafer and the InSb film, a sequence of buffer layers were grown as described elsewhere[18, 19]. First, a transition from GaAs to GaSb was produced, followed by InGaSb/InSb and InAlSb/InSb superlattices. Finally, the 1.5 μm InSb active later was grown on a thick InAlSb virtue substrate.

The room temperature carrier density and electron mobility of the InSb film before device fabrication were extracted via longitudinal and transverse resistance measurements in the van der Pauw (vdP) configuration, and were determined to be 3.34 ×$10^{12}$ cm$^{−2}$ and 57,000 cm$^2$/Vs, respectively. A 320 nm layer of Si$_3$N$_4$ was grown at 300 °C on top of the InSb thin film via plasma-enhanced chemical vapor deposition to serve as a hard mask. UV lithography was used to define the geometry of the semiconductor mesas and the Si$_3$N$_4$ film was etched away from unwanted areas under a CF$_4$ plasma using inductively coupled plasma reactive ion etching. A second dry etching step under a Cl$_2$ plasma at 180 °C was used to etch through the InSb until the buffer layer was reached. The Ti/Au (50 nm/300 nm) shunt and contacts were deposited in a side-contact configuration using electron beam physical vapor deposition and lift-off. The Ti film acts as an adhesion layer and diffusion barrier[20] but also serves to reduce the Schottky barrier height between the metals and the semiconductor as its work function $\Phi_{Ti}$ = 4.33 eV[21] is much closer to that of InSb, $\Phi_{InSb}$ = 4.57 eV[22], than that of Au, $\Phi_{Au}$ = 5.30 eV[23]. The final EMR devices were 1 mm in diameter.

To study the effect of annealing on the contact quality, devices were annealed for 1 hour first at 100 °C after which their properties were measured. Then they were further annealed at 200 °C and their properties measured. This procedure continued with 300 °C and 400 °C. All annealing steps were done in an Ar atmosphere at ambient pressure. The contact resistivity, $\rho_c$, was then extracted after each annealing step by using the transfer line method (TLM).[24] To determine their magnetic field response, all EMR devices were tested at room temperature in four-point measurement configuration under externally applied magnetic fields of up to 2 T. In this measurement configuration, an adjacent pair of contacts is used as the current source and drain whereas the opposite pair are used to measure the non-local voltage. The resistance at a magnetic flux density $B$ is given by:

$$R(B) = \frac{V_{nl}}{I} \tag{1}$$

where $V_{nl}$ is the voltage measured across the terminals and $I$ is the injected current. The magnetoresistance is then defined as:

$$MR(B) = \frac{R(B) - R_0}{R_0} \tag{2}$$

where $R_0$ is the resistance at zero-field. Two-dimensional finite element analysis (FEA) models were used to simulate the performance of our EMR devices. Previous work has demonstrated that the simulation results show good agreement with experimental InSb EMR data.[25, 26] The simulations were done with COMSOL Multiphysics software, the details of which can be found in our previous publications.[8, 26, 27] Unless stated otherwise, the following parameters were used in our models to closely match that of our experimental devices: outer device radius $r_o$ = 1 mm, inner device radius $r_i$ = (13/16)$r_o$, semiconductor thickness $d_s$ = 1.5 µm, semiconductor mobility $\mu_s$ = 50,000 cm$^2$/Vs, semiconductor carrier density $n_s$ = 2.0 ×10$^{16}$ cm$^{-3}$, metal thickness $d_m$ = 350 nm, metal mobility $\mu_m$ = 47.8 cm$^2$/Vs, and metal carrier density $n_m$ = 5.9 ×10$^{22}$ cm$^{-3}$. The conductivity, $\sigma_m$, of the metal was set to $\sigma_m = \mu_m n_m e \cdot (d_m/d_s)$ where $e$ is the electron charge and $d_m/d_s$ is a scaling factor to compensate for the fact that the 2-dimensional model assumes a uniform thickness across the entire device whereas experimentally the thickness of the metal (350 nm) differs significantly from that of the semiconductor (1.5 µm). The 2D model with a compensated metal conductivity was compared to a 3D model with accurate layer thicknesses and produced similar results (within 10%) at $B$ = 0 T, with the difference in the two models disappearing as the magnetic field was increased (see Supplementary Information). The 2D model was therefore chosen for further simulations due to its speed and simplicity. An interface resistivity, $\rho_c$, was also applied to the boundary between the metal and the semiconductor.

## 3  Results and Discussion

Transmission line measurements on the as-fabricated devices yielded an average $\rho_c$ of 5.9 ×$10^{-4}$ $\Omega cm^2$. According to literature, this value is well above the limit in which the contact resistivity begins to significantly dampen the EMR effect[7, 10] and thus thermal annealing was required to improve the contact quality. We investigated the effect of annealing on three types of devices: 1) van der Pauw discs of InSb to determine the carrier density, mobility, number of electronic bands, and intrinsic magnetoresistance[28]; 2) transmission line measurements to probe the electrical contact resistance and transfer length[24]; and 3) concentric circular EMR devices which resemble vdP discs with a metallic shunt to study extraordinary magnetoresistance response.

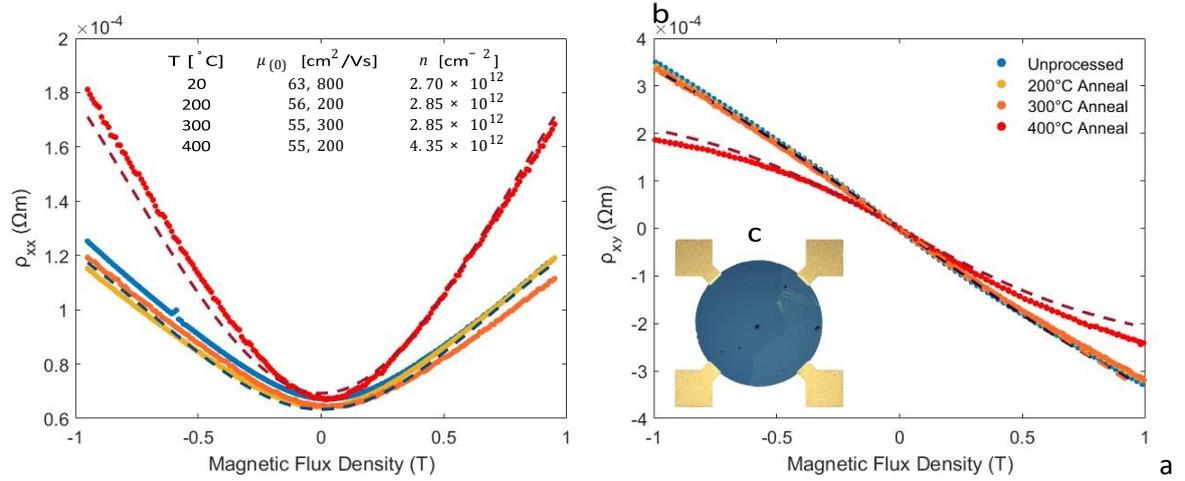

Figure 1: Magnetotransport measurements performed on a vdP disc of InSb showing the (a) $\rho_{xx}$ and (b) $\rho_{xy}$ signals after annealing for 1 hr at various temperatures. Solid circles represent experimental data points, whereas dashed lines are fits derived from the analytical expressions for two band conduction. (c) Optical microscope image of the vdP disc used to extract experimental data.

Figure 1 shows the longitudinal ($\rho_{xx}$) and Hall ($\rho_{xy}$) resistivity for a vdP disc of InSb and their field dependencies before and after thermal treatment. In the table inset of Figure 1, we list the mobility and carrier density values extracted from the vdP disc assuming that only one carrier band is active.[28] For the case of the unannealed sample, we observe an increase in $\rho_{xx}$, of 85% at 1 T and a nearly linear Hall resistivity, $\rho_{xy}$. This behavior remains largely unchanged after successive annealing steps at 200 °C and 300 °C. However, exposure to 400 °C results in a doubling of the intrinsic magnetoresistance to 170% and produces a strong non-linearity in the Hall signal. A simple one-band model can not explain the observation of an intrinsic magnetoresistance or a non-linear Hall resistivity, however features appear in the longitudinal and Hall resistivity curves which may yield some insights into the physical origins of the observed behavior, including: 1) At low $B$: MR $\propto B^2$, 2) at high $B$: MR $\propto B$, and 3) $\frac{d^2 \rho_{xy}}{dB^2} \neq 0$.

There are several theories which can explain the origins of an intrinsic linear magnetoresistance similar to what we observe in our samples at higher magnetic fields. The classical Parish-Littlewood model describes how linear non-saturating magnetoresistance can arise in strongly inhomogenous samples[29, 30], and it predicts that the cross-over from quadratic to linear MR should occur at $B_c = \mu^{-1}$. We can discard this explanation however as $B_c$ for our samples would be around 0.17 T, and we observe the linear transition around 0.6 T. The semiclassical guiding center diffusion (GCD) model predicts linear magnetoresistance for materials with a weakly varying disorder potential when the disorder correlation length is much larger than the cyclotron radius.[31] InSb fulfills many of the requirements for GCD: a high electron mobility, small chemical potential < 100 meV [32, 33], a high Fermi velocity $v_f \approx 1 \times 10^6$ m/s [34], and a moderate dielectric constant of 16.8 to screen Coulombic impurities[35] but we discard GDC as a potential explanation as it requires the Hall angle to be field independent (see Supplementary Information).

Linear magnetoresistance can also occur when approaching the extreme quantum limit of the density of states, i.e. when only the lowest Landau level is occupied, $k_B T \ll \hbar \omega_c$. The effective mass, $m^*$, of electrons in InSb is $(0.016 \pm 0.007)m_e$, where $m_e$ is the rest mass of an electron in vacuum.[36] At 1 T, this yields a cyclotron frequency, $\omega_c$, of 1.89 THz and $k_B T = \hbar \omega_c$ at T = 15 K. Our samples were tested at 300 K and are therefore far from the extreme quantum limit where this phenomenon can occur.

We propose that the origin of the intrinsic magnetoresistance in these InSb thin films can be best explained by a two-band conduction model. If multiple carrier species are present within a material, the longitudinal and Hall resistivities are given by the analytical expressions[37]:

$$\rho_{xx} = \frac{1}{e} \frac{(n_1\mu_1 + n_2\mu_2) + \mu_1\mu_2 B^2 (n_1\mu_2 + n_2\mu_1)}{(n_1\mu_1 + n_2\mu_2)^2 + \mu_2^1 \mu_1^2 B^2 (n_1 + n_2)^2}$$

$$\rho_{xy} = \frac{B}{e} \frac{(n_1\mu_1^2 + n_2\mu_2^2) + \mu_1^2\mu_2^2 B^2 (n_1 + n_2)}{(n_2\mu_2 - n_1\mu_1)^2 + \mu_1^2\mu_2^2 B^2 (n_1 + n_2)^2}$$

where $n_1$, $\mu_1$ and $n_2$, $\mu_2$ are the carrier density and mobility for the first and second bands respectively.

Here, the values of $n_1$, $n_2$, $\mu_1$, and $\mu_2$ were derived by fitting the analytical expressions for both $\rho_{xx}$ and $\rho_{xy}$ to the experimental data simultaneously with equal weighting. For the unprocessed sample, a good fit can be obtained if we assume the presence of two $n$-type bands: the first band with a carrier density of $2.25 \times 10^{12}$ cm$^{-2}$ and a carrier mobility of 60,000 cm$^2$/Vs, and the second band with a carrier density of $1.5 \times 10^{12}$ cm$^{-2}$ and a low mobility of 4,500 cm$^2$/Vs. We note that the carrier density and mobility in the highmobility band match well the values derived with the single-band model. The two-band model replicates the quadratic behavior at low fields and the linear transition that we observe in the experimental data. We propose that the second band of low mobility carriers are related to threading dislocations which form at the interface between the InSb film and the substrate[38], although there is also the possibility that low mobility species exist on the surface of the material.[3] Upon heating to 400 °C, the longitudinal resistivity at 1 T doubles and the Hall resistivity becomes strongly non-linear. This behavior can be well replicated by the two-band model if we assume a degradation of the mobility in both bands and a ten-fold increase in the density of the low mobility band: $n_1 = 2.475 \times 10^{12}$ cm$^{-2}$, $\mu_1 = 47,500$ cm$^2$/Vs, $n_2 = 1.35 \times 10^{13}$ cm$^{-2}$, and $\mu_2 = 1,300$ cm$^2$/Vs. The changes in electronic properties are likely due to the creation of secondary phases when InSb thermally decomposes above 350 °C as thermal excitations preferentially liberate antimony atoms from the surface of the material by forming a vapor, which leaves behind non-stochiometric InSb compounds.[39, 40]

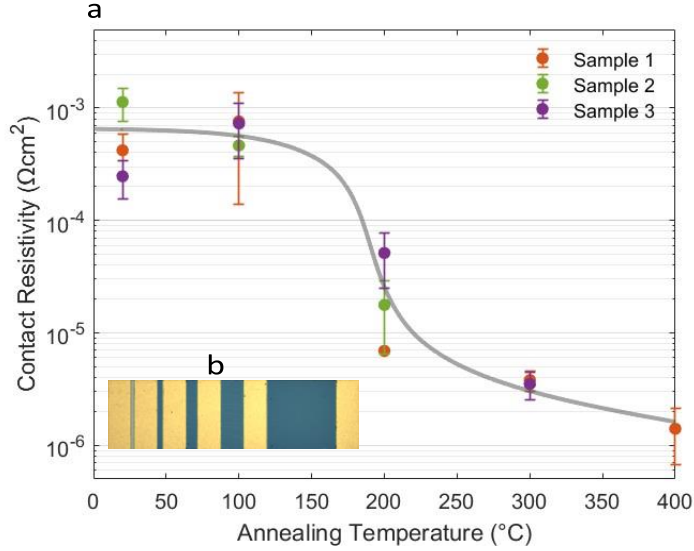

Figure 2: (a) $\rho_c$ of various InSb samples after annealing at various temperatures for 1 hr. (b) Optical image of TLM structure used to extract $\rho_c$.

We also examined the effect of annealing temperature on contact resistivity as shown in Figure 2. The as-fabricated samples showed a contact resistivity, $\rho_c$, of $5.9 \times 10^{-4}\,\Omega\mathrm{cm}^2$. Annealing at 100 °C in Ar under ambient pressure produced no clear change in $\rho_c$, but after exposure to 200 °C, the resistivity was decreased by an order of magnitude. An additional order of magnitude of improvement was realized after annealing at 300 °C, though further annealing at 400 °C yielded only a moderate improvement. We note that samples 2 and 3 were fabricated as side-contacted TLM structures whereas sample 1 had a top-contacted configuration. The similarity between the two types of contacts suggests that $\rho_c$ is relatively isotropic along the different crystallographic directions.

For a semiconductor with a given carrier concentration there is a lower limit to the contact resistivity due to the presence of a finite transmission probability, carrier velocity, and density of states. Baraskar et al. calculated the minimum $\rho_c$ in *n*-type InSb for carrier densities ranging from $10^{18}$ to $10^{21}$ cm$^{-3}$.[16] Extrapolating their calculations for the cases with low Schottky barriers to the carrier density extracted from our samples yields a minimum $\rho_c$ on the order of $1 \times 10^{-6}\,\Omega\mathrm{cm}^2$, similar to the saturated value we observe in our experiments of around $1.4 \times 10^{-6}\,\Omega\mathrm{cm}^2$. This indicates that we may be close to the lower limit for the value of the contact resistivity that can be achieved with low carrier densities in this particular material system, though further improvements might be realized by using InSb films with a highly doped contacting layer.[2]

EMR device performance as a function of annealing temperature is shown in Figures 3a and b. The device in Figure 3c was tested directly after fabrication before any annealing process had taken place and yielded a zero-field resistance, $R_0$, of 5.6 Ω and a modest MR of around 870% at 2 T. After annealing at 200 °C for 1 hr in Ar, two interesting features appear in the device response. $R_0$ is reduced to 1.9 Ω, as a direct result of the improved contact quality, but the minimum $R$ of 1.3 Ω is found not at $B = 0$ T but at a finite magnetic field of $0.14$ T. Additionally, the high-field resistance of the device becomes strongly asymmetric, yielding an MR of $1{,}220\%$ when the magnetic field is oriented in the positive direction and $2{,}450\%$ in the negative direction. Typically asymmetric responses in EMR devices are attained when the vertical symmetry of the geometry[26, 27, 41] or material properties[26] in a device are broken. Optical microscopy revealed no changes a    b in the shape of the sensor, indicating that geometric asymmetry could not explain our results. After annealing at 300 °C, $R_0$ drops by two orders of magnitude to 0.03 Ω, the MR increases to 65,000%, and the response of the device recovers its symmetry. However the high-field resistance is lower than that of the unannealed sample. The experimental data are compared with simulations in Figures 3a and b, which phenomenologically includes asymmetry in the material properties and variations in the contact resistance as described later. In the following, we discuss the main experimental features of a greatly enhanced MR, asymmetric device performance, and a decrease in the high field resistance with the discussion supported by numerical simulations.

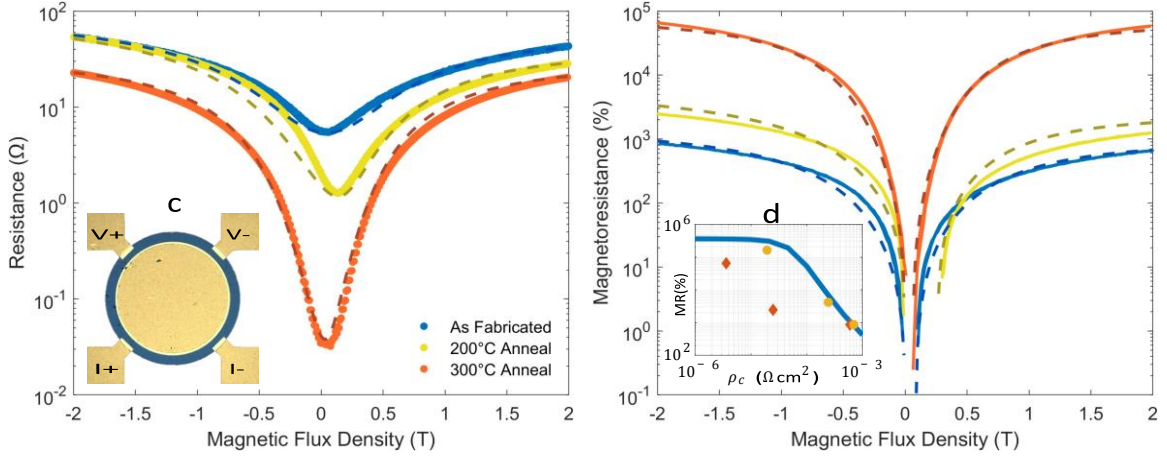

Figure 3: (a) Resistance as a function of magnetic flux density for the same device after annealing at various temperatures. Experimental data is shown as points whereas simulation results are displayed as dashed lines. (b) Calculated magnetoresistance values for data presented in Fig. 2a. (c) Optical image of the experimental device. (d) Magnetoresistance at 1 T as a function of contact resistivity for experimental data against the $\rho_c$ extracted from the TLM measurements (♦), extrapolated MR if the high field resistance remained constant against the best-fit $\rho_c$ from simulations (•), and simulations (-).

To help understand our results we consider the behavior of electronic current in an idealized EMR device. At zero magnetic field, the current in high-performing devices flows primarily through the metal shunt as depicted in Figure 4, which results in a low $R_0$. This current shunting is effective when the barrier for entering the metal is low, thus annealing leads to a large decrease in $R_0$ as the contact resistance is reduced by more than two orders of magnitude. At high fields, when the majority of the current passes through the semiconductor portion of the device, the resistance is primarily determined by the properties of the semiconductor. Therefore, if the properties of the semiconductor device remain unaffected by the annealing, the high-field resistance of our devices should stay fixed throughout the annealing process and the increase in magnetoresistance is then achieved solely by a decrease in $R_0$ as reported elsewhere.[8] However, our devices show a decrease in resistance at high fields suggesting a change in the properties of the semiconductor which may cause both the asymmetry observed in the 200 °C case and a tempering of the magnetoresistance. We believe that our results can be explained by the interdiffusion of Au and Ti and subsequent diffusion of Au into the InSb structures. We deposit Ti as a barrier layer, however Au and Ti interdiffusion can occur at temperatures as low as 200 °C and form an intermetallic phase[20]. Diffusion processes can be described using the Arrhenius equation:

$$D = D_0 e^{-\Delta E_a/k_B T} \tag{5}$$

where $D$ is the diffusivity, $D_0$ is the pre-exponential factor, $\Delta E_a$ is the activation energy, $k_B$ is the Boltzmann constant, and $T$ is the temperature. The diffusivity factors for Ti in Au are $D_0 = 5.0\times10^{-11}$ cm$^2$s$^{-1}$ and $\Delta E_a$ = 0.66 eV.[20] Once the Au atoms reach the InSb interface they can become highly mobile; the diffusivity factors for Au in InSb are $D_0 = 7.0\times10^{-4}$ cm$^2$s$^{-1}$ and $\Delta E_a$ = 0.32 eV.[42] Jany et al. showed with density functional theory (DFT) calculations how an AuIn$_2$ alloy is formed at Au/InSb interfaces as Au replaces atoms in the Sb sublattice. At low concentrations (≤20 at.%), their predictions also suggest that Au is primarily incorporated into the Sb sublattice, whereas at higher concentrations the gold atoms are equally present in both sublattices and tend to agglomerate, forming wirelike structures inside the semiconductor.[43] The formation of AuIn$_2$ intermetallic phases may increase the carrier density in the semiconductor and therefore decrease its overall resistivity. This interpretation is supported by experiments as shown in the Supplementary Information. This may also explain why we do not see a change in the behavior of the unshunted vdP disc (Figure 1) after annealing at 300 °C, as the metal-semiconductor contact area is small and only exists at the contacts.

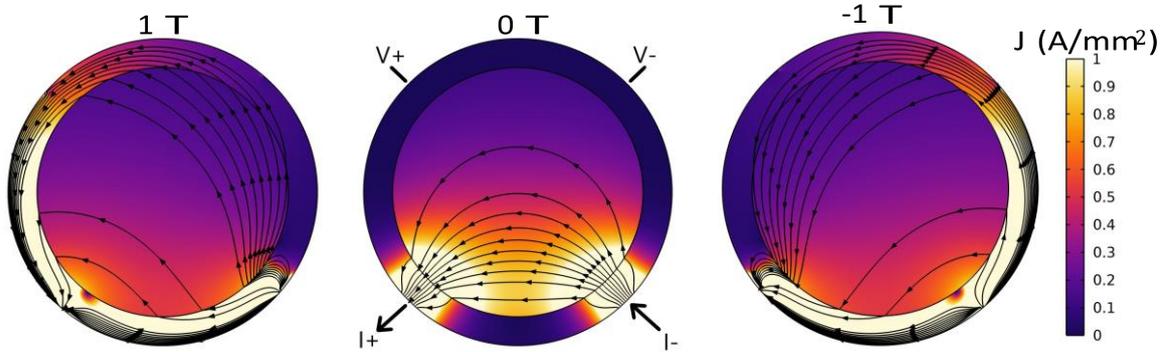

Figure 4: Simulations of the current density in an EMR device at B = 1 T, 0 T, and -1 T.

To gain further understanding of the behavior of our devices, FEA simulations were done to try and replicate the experimental results. The numerical and experimental data yielded a good agreement, as can be seen in Figures 3a and b. In the following, we explain how the simulations were used to elucidate the observed behavior of our EMR device. Figure 4 illustrates how the current density is distributed inside an EMR device at 0 and ±1 T. In the absence of an applied magnetic field, the injected current travels primarily through the gold shunt in the middle of the device and passes through the semiconductor only in the regions closest to the contacts. The resistance of the device is therefore primarily determined by the properties of the shunt ($\rho_{Au}, \rho_c$). When a strong magnetic field is applied, the current is forced to travel through the semiconductor instead. The region between the current contacts has a high current density regardless of the direction of the applied magnetic field. The field directionality influences whether the remaining current travels into the shunt directly before being redirected into the upper part of the semiconductor, or through the semiconductor near the injection point before entering the upper part of the shunt and exiting in the region closest to the drain. If the material is homogeneous, then the two sides are equivalent and the device response is symmetric. However, it becomes clear that in an inhomogenous device the semiconductor on one side of the disc should have a stronger influence on the potential drop across the voltage contacts when the field is oriented in one direction, and the other side should dominate the response in the opposite field direction.

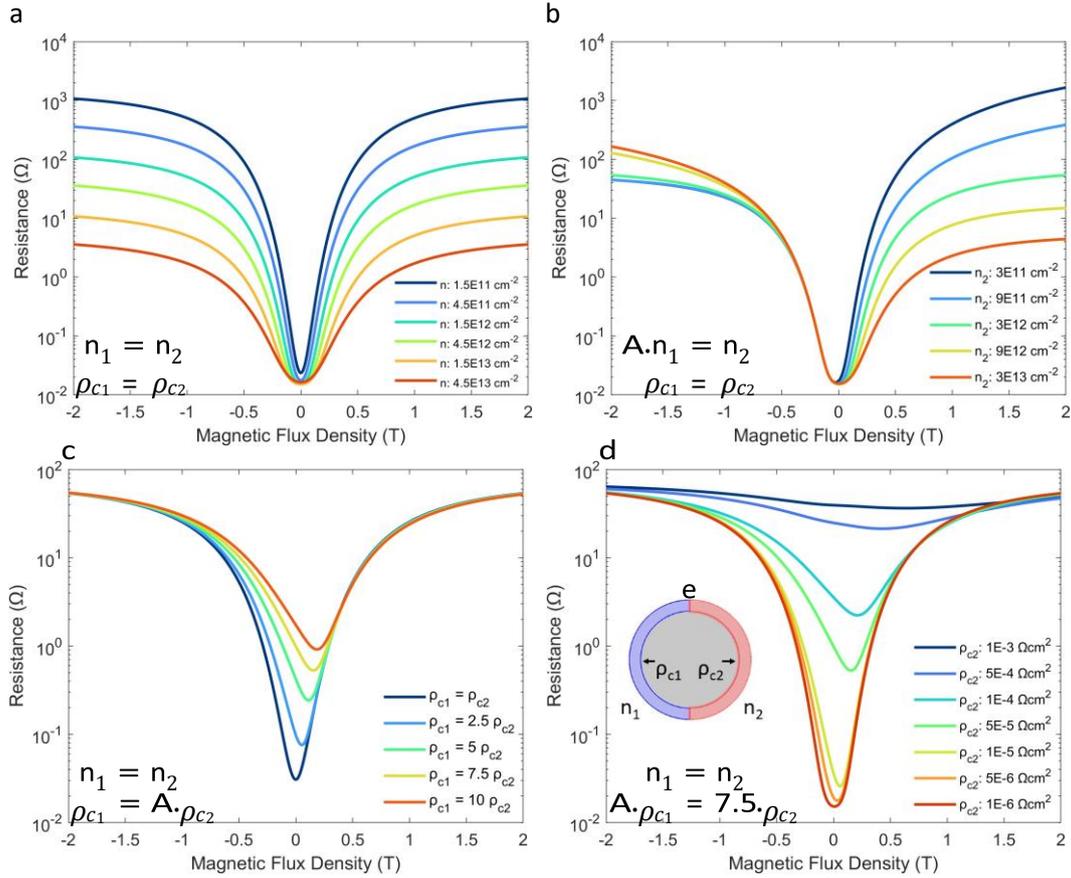

Figure 5: Effect of semiconductor properties ($n_1$, $n_2$, $\rho_{c1}$, $\rho_{c2}$) on the resistance of EMR devices as a function of magnetic flux density for four different cases: (a) uniformly varying the carrier density with a uniform $\rho_c = 1 \times 10^{-6}$ Ωcm², (b) varying $n_2$ with a fixed $n_1$ and a uniform $\rho_c = 1 \times 10^{-6}$ Ωcm², (c) varying $\rho_{c1}$ with a fixed $\rho_{c2} = 5 \times 10^{-5}$ Ωcm² and a uniform carrier density $n = 3 \times 10^{12}$ cm⁻², and (d) varying $\rho_{c2}$ with a fixed ratio of $\rho_{c1}/\rho_{c2} = 7.5$ and a uniform carrier density $n = 3 \times 10^{12}$ cm⁻². The geometry of the device used in the FEA simulations is shown in inset (e).

We begin our simulation study by examining the influence of uniformly varying the semiconductor carrier density on the resistance of the EMR sensor, as shown in Figure 5a. The high-field resistance of the device is inversely proportional to the carrier density, whereas the zero-field resistance remains largely unchanged. The highest resistances observed in our devices prior to annealing can be matched by the simulations if we assume a carrier density of $3.15 \times 10^{12}$ cm⁻², a value very similar to the experimentally extracted carrier density of $3.30 \times 10^{12}$ cm⁻². If the properties of the semiconductor are uniform, then the resistance is symmetric around B = 0 T. To break this symmetry, we split the semiconductor portion into two sections along the vertical axis of the device, as shown in Figure 5e.

Figure 5b shows the effect of varying the carrier density of only one side of the split semiconductor. The high-field resistance is significantly affected but only in the positive magnetic field direction, with the resistance in the negative field direction remaining largely the same. It can be seen that the material properties of the semiconductor on each side of the line of vertical symmetry determine the resistance for one of the magnetic field orientations. However, in these cases, the minimum resistance of the device can still be found at B = 0 T;

Table 1: Overview of Simulation Parameters

| T [°C] | $\mu$ [cm$^2$/Vs] | $n_1$ [cm$^{-3}$] | $n_2$ [cm$^{-3}$] | $\rho_{c1}$ [Ωcm$^2$] | $\rho_{c2}$ [Ωcm$^2$] | $\rho_{c1}/\rho_{c2}$ |
|---|---|---|---|---|---|---|
| 20 | $5 \times 10^4$ | $2.10 \times 10^{16}$ | $2.31 \times 10^{16}$ | $7.12 \times 10^{-4}$ | $5.70 \times 10^{-4}$ | 1.09 |
| 200 | $5 \times 10^4$ | $2.55 \times 10^{16}$ | $3.11 \times 10^{16}$ | $1.20 \times 10^{-4}$ | $3.69 \times 10^{-5}$ | 3.25 |
| 300 | $5 \times 10^4$ | $4.85 \times 10^{16}$ | $5.00 \times 10^{16}$ | $2.50 \times 10^{-5}$ | $1.43 \times 10^{-5}$ | 1.75 |

this form of symmetry breaking does not seem to affect the minimum resistance point. In Figure 5c, we show the effect of another type of asymmetry by applying a different value of $\rho_c$ to each side of the device. Here the carrier density is set to be uniform across the device and the value of the contact resistivity on one side, $\rho_{c1}$, is varied while the other side, $\rho_{c2}$, is fixed. Increasing the difference between the two resistivities shifts the point of minimum resistance to a finite value away from $B = 0$, with the magnitude of the shift depending on the ratio $\rho_{c2}/\rho_{c1}$. In Figure 5d, we show the effect of a fixed ratio of $\rho_{c2}/\rho_{c1}$ when the magnitude of the resistivity is varied. It can be seen that an inhomogenous contact only affects the resistance curve if the contact resistivity is above a threshold value. At very low resistivities, the resistance is symmetric around $B = 0$, but the more the contact resistivity is increased, the higher the threshold value and the asymmetric shift become. Our experimental data for the 200 °C case in particular can therefore be explained by two mechanisms: 1) the presence of an inhomogenous contact resistance along the perimeter of the metal shunt with a sufficient magnitude to be sensitive to this asymmetry and 2) an unequal diffusion of Au into each side of the device producing regions with different carrier densities. In Table 1 we show the physical property parameters used in the simulations from Figure 3a and b in order to match the experimental data.

However, the divergence between the experimental data and the model seen in Figure 3d is not completely explained, even when accounting for the drop in the high-field resistance, as it would only improve magnetoresistance by a factor of 2. It is also possible that the contact resistivity in the EMR device is higher than what was measured in the TLM structures, and thus the experimental data points should be shifted further right. For the unannealed sample with the highest contact resistivity, the simulations show good agreement with the experiment, it is only after annealing where we do not know the contact resistivity of the EMR device itself that we start to see a difference. In the same figure, using yellow circles, we plot the magnetoresistance as if the high-field resistance had remained constant during the annealing steps against the average value of $\rho_c$ which the produced the best fit in the simulations (see Table 1). Here we reach a much closer agreement with the predicted magnetoresistance suggesting that the devices may have a higher contact resistivity than the TLM structures and that the performance could also be improved if one could ensure that the high-field resistance did not decrease after annealing.

## 4 Conclusion

We investigated the effect of annealing on the contact resistivity, $\rho_c$, in InSb EMR devices with Ti/Au contacts and found that $\rho_c$ could be reduced by two orders of magnitude by annealing at 300 °C. Exposure to higher temperatures resulted in degradation of the semiconductor material. We also examined the unannealed intrinsicmagnetoresistance of InSb thin films and determined the most likely mechanism behind the magnetoresistance is the presence of multiple conduction bands due to the existence of defects at the substrate interface and/or surface. The multiple band effect becomes more pronounced in the Hall resistance after exposure to high annealing temperatures as additional low-mobility carriers are generated. EMR devices were tested before and after various annealing steps, and the device performance was found to significantly improve as $\rho_c$ was decreased. We also demonstrated how annealing can produce inhomogenous device properties which lead to an asymmetric device response.

# 5 Acknowledgements

The authors acknowledges the support of Novo Nordisk Foundation Challenge Programme 2021: Smart Nanomaterials for Applications in Life-Science, BIOMAG Grant NNF21OC0066526. D.V.C. and T.S. also acknowledge the support of Novo Nordisk Foundation NERD Programme: New Exploratory Research and Discovery, Superior Grant NNF21OC0068015.

# Supplementary Information

The Hall angle, $\theta_H$, of the InSb samples before and after thermal treatment was calculated using the expression[31]:

$$\tan^{-1}(\theta_H) = \frac{\rho_{xy}}{\rho_{xx}} \tag{6}$$

Figure 6 shows the Hall angle and intrinsic magnetoresistance of the samples as a function of $\mu B$. The Hall angle varies as a function of $\mu B$ for all of the samples tested, although it is lower for samples which were annealed at 400 °C. The magnetoresistance shows two regimes: superlinear below $\mu B \approx 1$, and linear above it.

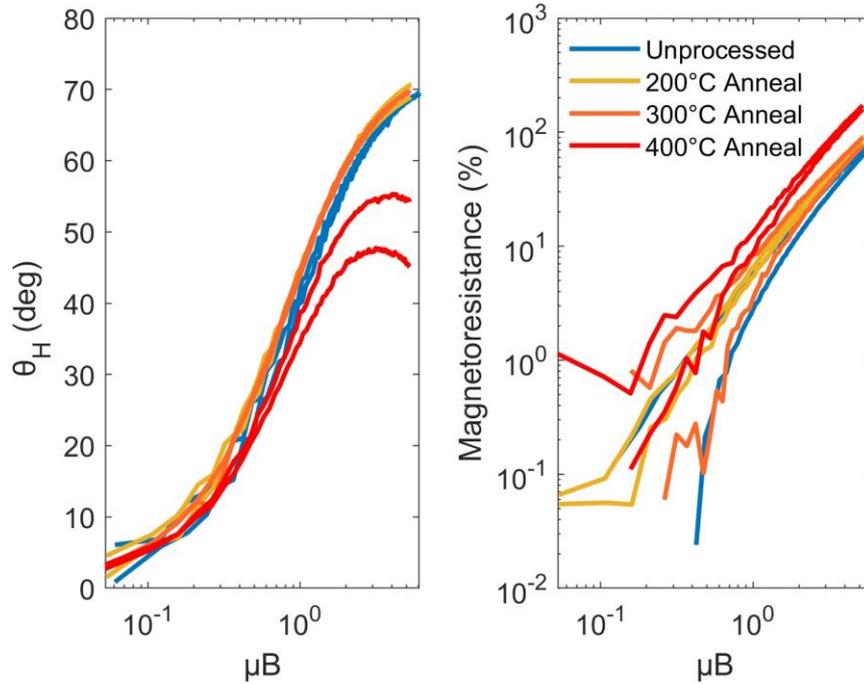

Figure 6: Intrinsic magnetoresistance of InSb samples before and after thermal annealing. The Hall angle is plotted as a function of $\mu B$ (left). Magnetoresistance vs. $\mu B$ on a logarithmic scale (right).

After annealing at 300 °C and performing a focused ion beam cut along the contact region, the cross-section of the InSb contacts was analyzed by scanning electron microscopy. Figure 7 shows the interface between metal and semiconductor at two different contact areas. The dry etching process produces an undercut beneath the $Si_3N_4$ hard mask. The hard mask creates a shadowing effect over the majority of the sidewall resulting in only a narrow point-contact at the base of the semiconductor (Figure 7a). Additionally, the Ti layer is intended to function as a diffusion barrier, but since the sidewall is only partially covered, Au can be found in direct contact with the semiconductor, allowing for rapid diffusion of metal into the semiconductor. In some contact regions it was observed that the hard mask had bent, decreasing the shadowing effect and increasing the amount of semiconductor contacted by metal (Figure 7b). This variation in sidewall coverage could also explain part of the inhomogeneity in the contact resistance along the device perimeter.

Analysis of our experimental data suggests that the carrier concentration in the semiconductor increased as a result of annealing. In order to determine whether this could be explained by the diffusion of Au into the device, electron diffraction spectroscopy (EDS) was used to quantify the atomic percentage of the various elements across a contact point, as shown in Figure 8. While the Sb signal decays very quickly and remains low and constant on the other side of the interface, the Au signal decays slowly and extends deeply into the semiconductor region, suggesting that Au diffusion has occurred.

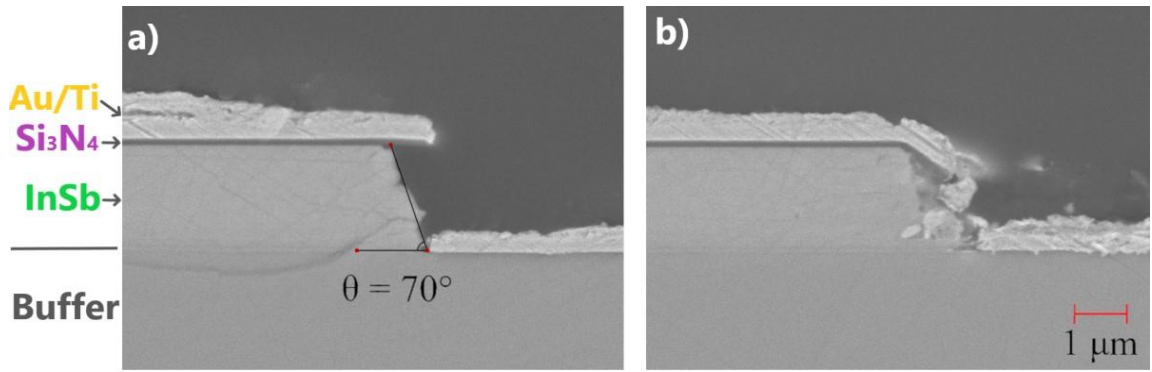

Figure 7: Scanning electron micrographs of the cross-section of an annealed device showing the contact region between metal and semiconductor at two different points. The different materials are indicated on the left.

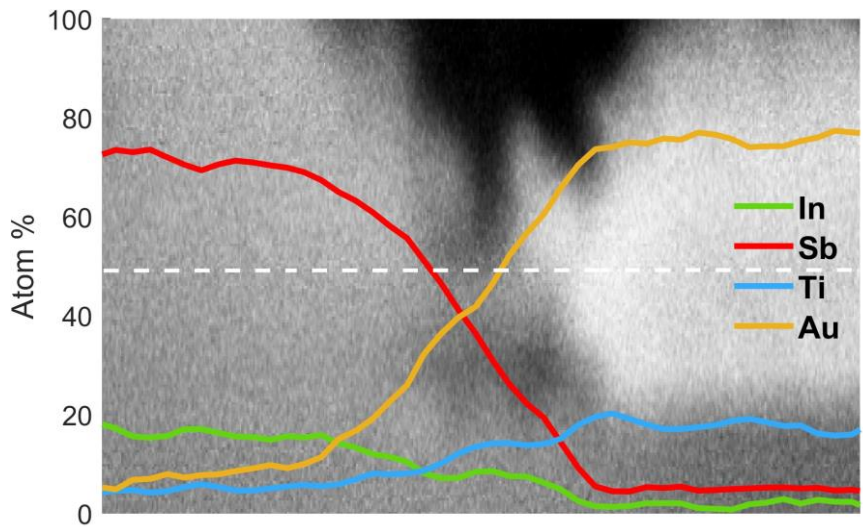

Figure 8: Results of electron diffraction spectroscopy showing the atomic percentages of different atoms superimposed on a scanning electron micrograph of the studied region. The semiconductor region appears in grey on the left while the metal region is the brighter material on the right.